# An Approach Towards Physics Informed Lung Ultrasound Image Scoring Neural Network for Diagnostic Assistance in COVID-19

Mahesh Raveendranatha Panicker, Yale Tung Chen, Gayathri M, Madhavanunni A N, Kiran Vishnu Narayan, C Kesavadas and A P Vinod

*Abstract*— Ultrasound is fast becoming an inevitable diagnostic tool for regular and continuous monitoring of the lung with the recent outbreak of COVID-19. In this work, a novel approach is presented to extract acoustic propagation-based features to automatically highlight the region below pleura, which is an important landmark in lung ultrasound (LUS). Subsequently, a multichannel input formed by using the acoustic physics-based feature maps is fused to train a neural network, referred to as LUSNet, to classify the LUS images into five classes of varying severity of lung infection to track the progression of COVID-19. In order to ensure that the proposed approach is agnostic to the type of acquisition, the LUSNet, which consists of a U-net architecture is trained in an unsupervised manner with the acoustic feature maps to ensure that the encoder-decoder architecture is learning features in the pleural region of interest. A novel combination of the U-net output and the U-net encoder output is employed for the classification of severity of infection in the lung. A detailed analysis of the proposed approach on LUS images over the infection to full recovery period of ten confirmed COVID-19 subjects shows an average five-fold cross-validation accuracy, sensitivity, and specificity of 97%, 93%, and 98% respectively over 5000 frames of COVID-19 videos. The analysis also shows that, when the input dataset is limited and diverse as in the case of COVID-19 pandemic, an aided effort of combining acoustic propagation-based features along with the gray scale images, as proposed in this work, improves the performance of the neural network significantly and also aids the labelling and triaging process.

I. INTRODUCTION

LUNG ultrasound (LUS) has been one of the most promising clinical tools for the diagnosis and monitoring of various pulmonary diseases [1]. LUS typically uses a pulse-echo principle to image the lungs where an acoustic pulse is sent out to the lung tissue from a transducer and the reflected echoes are received by the same transducer, which is further processed to obtain a diagnosable image [2]. The extent of reflection depends on the acoustic impedance at the different interfaces within the lung. A healthy lung is defined by a balanced ratio of tissue, air, and fluid. The large difference in the acoustic impedance between the visceral pleura (the membrane that covers the surface of each lung) and the air inside the alveolar space, gets manifested as specular reflector that completely reflects the ultrasound (US) waves thus obstructing the penetration into further depths. This makes the LUS examination more challenging as further investigation of the regions beneath the pleura is difficult. However, the high-energy reflected US waves get reflected multiple times between the pleura and the transducer inducing reverberations [3]. The reverberations repeatedly appear as hyper-echoic horizontal lines of decreasing intensity beneath and parallel to the pleural line at equidistant intervals. The pattern is called the A-line (refer to class 1 in Fig. 1) which is a clinical representation of a healthy lung [3]. However, the presence of pulmonary diseases would alter the specular behavior of the pleura deterring the A-lines and inducing newer acoustic artifacts such as B-lines (refer to class 3 in Fig. 1) or C-lines (refer to class 5 in Fig. 1) depending on the severity [4]. The aforementioned prospect of easy demarcation of a healthy and a diseased lung by the visual assessment of the acoustic artifacts give LUS an upper hand over other imaging modalities [2]. However, the accuracy of diagnosis is strongly influenced by the localization and interpretation of the artifact profiles [5].

Recently, LUS has been gaining wide traction with the spread of novel Corona Virus Disease 2019 (COVID-19). The virus potentially affects the lungs leading to acute respiratory distress syndrome (ARDS), pneumonia, acute dyspnea, and hypoxemia [6]. Interestingly, the pathological conditions of the lung associated with COVID-19 could be majorly characterized by imaging the pleura or the pleural space and the associated artifacts. The pulmonary manifestations are reported in the form of pleural thickening, focal, multifocal, and confluent B-lines,

This work was supported in part by the Department of Science and Technology - Science and Engineering Research Board (DSTSERB (CVD/2020/000221)) and the Corporate Social Responsibility (CSR) funding from Federal Bank.
Mahesh Raveendranatha Panicker, Gayathri M, Madhavanunni A N and A P Vinod are with Center for Computational Imaging, Indian Institute of Technology, Palakkad, India (e-mail: mahesh@iitpkd.ac.in).
Dr. Yale Tung Chen is with Department of Emergency Medicine, Hospital Universitario Puerta de Hierro, Spain.
Dr. Kiran Vishnu Narayan is with Department of Pulmonary medicine, Government Medical College, Thiruvananthapuram, India.
Dr. C Kesavadas is with Imaging Sciences and Interventional Radiology, Sree Chitra Institute of Medical Sciences and Technology (SCTIMST), Thiruvananthapuram, India.

lung consolidations and re-appearance of A-lines during recovery [7]. Recent studies also report that LUS outperforms standard chest radiography and yields results similar to CT in the monitoring of pneumonia and ARDS associated with COVID-19.

With COVID-19 outbreak, there has been a resurgence in automatic classification and scoring of lung ultrasound images to determine the severity of lung anomalies [8, 9]. In [10, 11], a convolutional neural network (CNN) based algorithm is employed for automatic detection of B-lines, merged B-lines, lack of lung sliding and pleural consolidation. However, the approach was trained on in vivo swine models and simulated M-mode models. In [12], a novel approach which will help the clinicians in tracking the progress of the COVID-19 infection in subjects, particularly in emergency medicine has been proposed. However, in the case of data-driven approaches as in [10, 11, 12], the decisions are based on the training datasets that are categorized by scores/labels indicating the severity. This creates bias as the scoring/labelling procedures could be subjective and may not be accurate [13, 14]. Moreover, this is also affected by the acquisition system and the settings with which the data is captured which may or may not result in a good quality LUS image. Therefore, there is a necessity for an exhaustive training to avoid ambiguous and unpredicted outcomes [15]. This is a challenging task in the present situation of COVID-19 due to the limited availability of datasets, dedicated clinicians who can complement the scoring/labelling, privacy, transparency and legal issues, constraints in time and non-deterministic progression of the virus disease [13-16].

The proposed approach tries to address the three main concerns 1) to reduce the bias in scoring/labelling process and to aid the clinicians in aided annotations by enhancing the image to save their time, 2) to reduce the dependence on the type of data acquisition and 3) to ensure that the neural network is learning the region below pleura and characteristics such as A and B lines. It is important that any effort towards applying neural networks for COVID-19 LUS images should be an aided effort rather than relying completely on the raw image input or image augmentation approaches [12]. By incorporating image rectification before further processing of images, irrespective of the acquisition using linear or convex or sector probes make the images agnostic to the type of probe. An image processing algorithm aligned with US physics to characterize the region below pleura, uniquely represented by features such as local phase map, acoustic shadow map, and integrated back scatterer energy are employed to train a neural network, named as LUSNet, to ensure that the LUSNet is learning the region below pleura and characteristics such as A and B lines. A study on improving the labelling process by reducing differences among the annotations using the feature maps along with the images for annotations is also presented. The rest of the paper is organized as follows. Section II discusses the proposed approach in detail. The detailed analysis and results in various LUS conditions are presented in Section III. In Section IV, a detailed analysis of various COVID-19 stages over infection to full recovery period of two subjects is presented. Section V concludes the paper with possible directions for future work.

## II. MATERIALS AND METHOD

The severity classes of COVID-19 are decided from lung manifestations during disease progression as discussed in [8, 9]. It includes the presence of A-lines (Class1 in Fig. 1), reduction in the pleural thickness, lack of A-lines (Class 2 in Fig. 1), appearance of single or multiple B-lines (Class3 in Fig. 1) to a confluent appearance of B-lines (Class4 in Fig. 1) and further degraded by the appearance of C-lines (Class5 in Fig. 1) due to the effusion in between two pleural surfaces with or without the appearance of air bronchograms. In this work, the COVID-19 images are classified into above mentioned five classes as observed by the clinicians during the progress of the disease in ten subjects over 21 days. The various class images shown are taken from different subjects to give a good representative image of the five target classes. To aid quick triage and diagnosis, a novel approach is proposed in this paper, where the first step is rectification of the ultrasound images to make it agnostic to the type of probe employed and to restrict unwanted edge effect particularly in the case of convex and sector probes. The second step is the generation of ultrasound propagation driven feature maps such as shadow map, integrated back scatter energy map and local phase map. As part of the third step, a fused image is formed by summing the original image, the local phase image and the image of the product of shadow map and integrated back scatterer energy, which aids in the quick decision making of the clinician irrespective of the quality of image acquisition as well as reduce the bias in the labelling/scoring process. The fourth step in the proposed approach deploys the neural network based on the popular U-net architecture [17] for classifying the input into one of the five severity classes. The encoder-decoder architecture learns to minimize the distance between the fused image (ground truth)

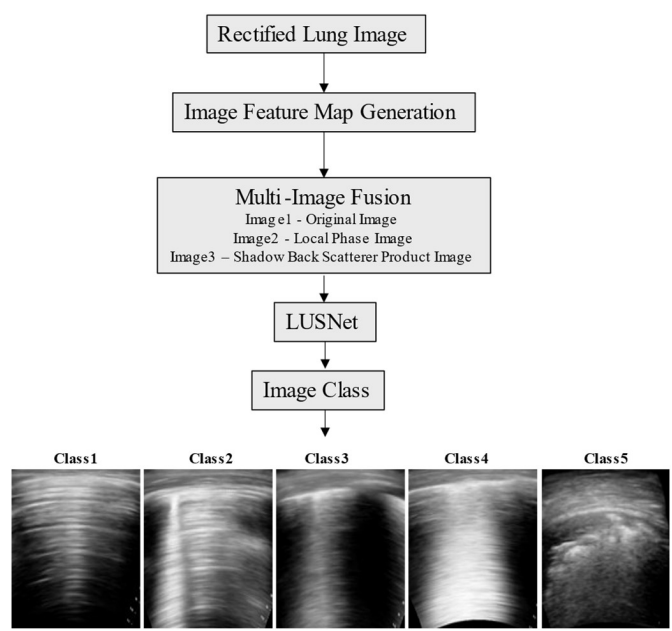

Fig. 1. Proposed LUSNet approach

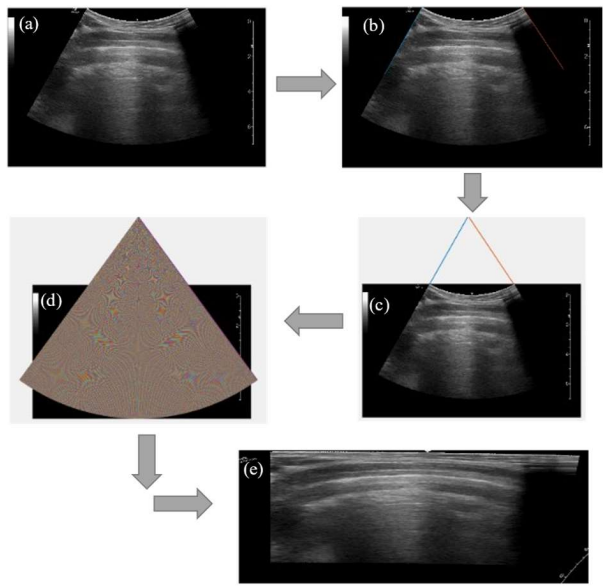

Fig. 2. Steps in image rectification algorithm. (a) original image, (b) manual selection of the edges of the image region, (c) automatic identification of the origin of convex image, (d) grid generation, (e) linear mapping of images

and the original image so that the proposed network is forced to learn LUS features, while trained in supervised fashion to classify the images into five classes. The algorithmic steps of the proposed approach are as shown in Fig. 1 and explained in detail in the following subsections.

*A. Image Rectification and Generation of Image Features*

As discussed before, all the automatic classification and scoring of lung ultrasound images proposed in the literature so far [10-12] are based on or tuned to certain acquisition setups and systems. The proposed approach is an attempt towards reducing this dependence in the case of COVID-19 diagnostics. Towards this, the first step is removing the effect of the type of the probe (particularly the edge artefacts in the convex and sector probes) through the conventional rectification approach as shown in Fig. 2. Once the images are rectified, the next step is to generate the ultrasound propagation driven feature maps.

The feature maps employed in this work are motivated by the bone segmentation works in [18], as pleura is similar to bone in terms of its specular characteristics [19]. The selection of the feature maps is done based on two aspects 1) they should highlight the pleura, A and B lines and 2) they should have higher energy below the pleural region which is the region of interest. Accordingly, the features of local phase image (LPI), pleura shadow (SH) image and integrated back scatter (IBS) image are chosen for further processing.

For generating the LPI, the rectified image is squared twice so that the high intensity reflections are enhanced before further processing. The local phase image is obtained using [18] as below (1):

$$LPI(x,y) = 1 - arctan\left(\frac{\sqrt{m_2(x,y)^2 + m_3(x,y)^2}}{m_1(x,y)}\right) \quad (1)$$

where, $m_1$, $m_2$ and $m_3$ are the three different components of monogenic signal image (specifically $m_1$ being the real part of the even component, $m_2$ being the real part of the odd component and $m_3$ being the imaginary part of the odd component respectively) calculated using Reisz filter employing a log Gabor band pass filter [20] for each of the pixel coordinates $(x,y)$.

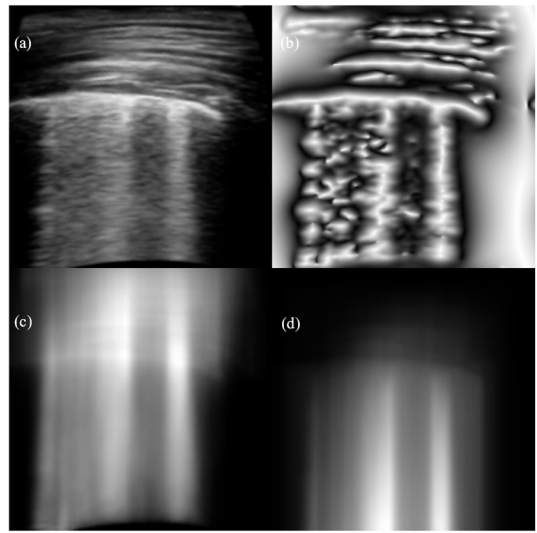

Fig. 3. Image feature generation. (a) rectified original image, (b) local phase image (c) shadow image and (d) integrated back scatter image

The IBS image and SH image will help in identifying the region below pleura due to the presence of A and B lines (or patch) when compared to soft tissue above pleura. The IBS is calculated as in (2), where $IBS(x,y)$ represents the IBS image for each pixel $(x,y)$ and $I$ is the per pixel intensity.

$$IBS(x,y) = \sum_{k=1}^{x} I^2(k,y) \quad (2)$$

The shadow value is calculated for each pixel as Gaussian weighted accumulation of the pixels below it, as per (3).

$$SH(x,y) = \frac{\sum_{k=x}^{R} G(k-x)I(x,y)}{\sum_{k=x}^{R} G(k-x)} \quad (3)$$

In (3), $SH(x,y)$ is the shadow value of a pixel at row, $x$ and column, $y$. $G$ represents the 1-D Gaussian weighting function of size $R$ and standard deviation of $R = 4$, where $R$ is the number of rows in the image and $I$ represents the image pixel intensity at row, $x$ and column, $y$. An illustrative example taken from the COVID-19 case with Class3 (scattered B-lines) is shown in Fig. 3 to show the LPI, IBS and SH along with the rectified original image. It is very evident from Fig. 3 that it is possible to highlight the pleural region by combining the images (a to d), which will be discussed in next section.

*B. Multi-Image Fusion*

It is important to note that the LPI, SH image and IBS image features are by themselves capable of highlighting the A, B, or C lines and thus could be potentially used for classifying the image into one of the five classes. As can be seen from SH and IBS images of Fig. 3, the characterization of the B-line is very evident. A dot product between SH and IBS images as shown in the third column in Fig. 4 clearly highlights the region below

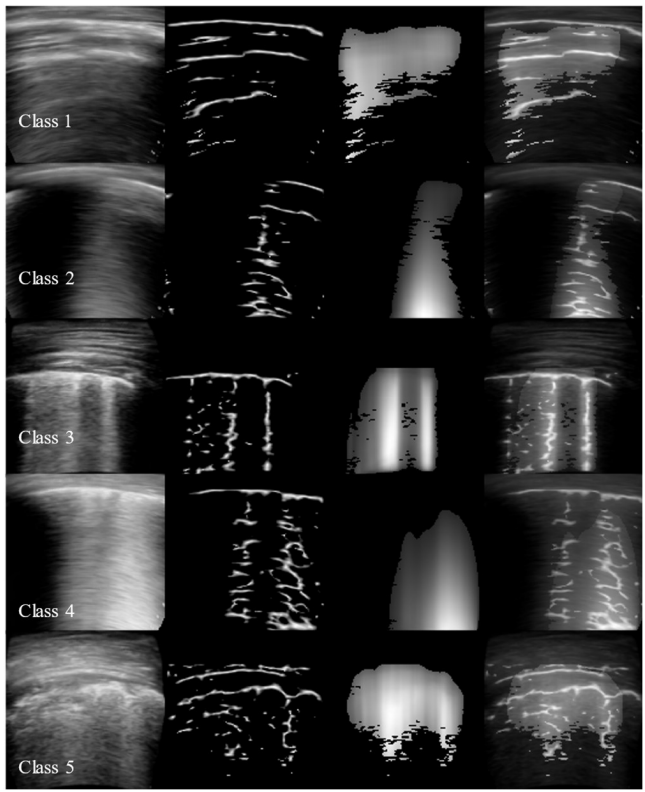

Fig. 4. Images and features maps across the five classes. In column1: Original Image, column2: Local Phase Image, column3: SHIBS Image, column4: fused image (sum all the image and feature maps)

pleura. For example, comparing the rows in Fig. 4, it is evident that, with the reduction in the pleural thickness and increase in lung consolidations, the B-lines have become more confluent and are picked up clearly by the IBS × SH image. Thus, instead of using the SH and IBS as individual images, it has been used in combination as their dot product referred to as the shadow IBS (SHIBS) image in this work. A detailed analysis of the images and the feature map for various classes is as shown in Fig. 4. The first column are the rectified images, the second, third and fourth columns are the LPI, SHIBS and fused image which is the sum of original image, LPI and SHIBS images. The ability of fused images in clearly highlighting the pleural region of interest is extremely useful for aiding the clinicians in easier annotation as well as in unsupervised training of the U-net encoder-decoder architecture in the proposed LUSNet.

C. Proposed LUSNet

An overview of the proposed LUSNet is as shown in Fig. 5. The architecture is based on the U-net architecture [17] and the encoder-decoder backbone of the U-net was experimented with Resnet34 [21], VGG16 [22] and Inception V3 [23]. But the results show that all the methods perform in a similar fashion, with inceptionV3 being slightly superior. The proposed architecture has two parts, 1) the conventional U-net segmentation section (where instead of the standard sigmoid activation layer, a rectified linear unit (ReLU) layer is used) which learns in unsupervised manner to reduce the distance between the output and the fused image presented in section II.B and 2) the classification architecture which concatenates the bottleneck layer output (of U-net) and segmentation (decoder) output after passing through a post processing network. The first part ensures that the architecture is learning features relevant to the pleural region of interest. The architecture details of the post processing networks are as given in Table I.

As discussed, the model is trained simultaneously for minimizing the mean square error (MSE) between ReLU activated U-net decoder output, $Y$ and the fused image output, $X$ (orange dotted line as shown in Fig. 5) as well as for the categorical cross-entropy loss function on the softmax outputs, $\hat{y}_i$, against the clinical annotations, $y_i$ with $C = 5$ classes. The total loss function of the LUSNet is as shown in (4).

$$loss = \lambda_1 \|X - Y\|_2 - \lambda_2 \sum_{i=1}^{C} y_i \log(\hat{y}_i) \qquad (4)$$

where, $\lambda_1$ and $\lambda_2$ are the hyper parameter weights for each of the loss function, chosen to be 0.3 and 0.7, respectively. The LUSNet is optimized with Adam [24] with an initial learning rate of 0.0001. A dataset comprising of 5000 images, with almost equal distribution of the five classes is used for training. A dropout of 0.5 is employed for dense layer [25]. The LUSNet is built with Keras environment in Python and trained on an HP Z4 G4 workstation which has an 8GB P4000 NVIDIA GPU.

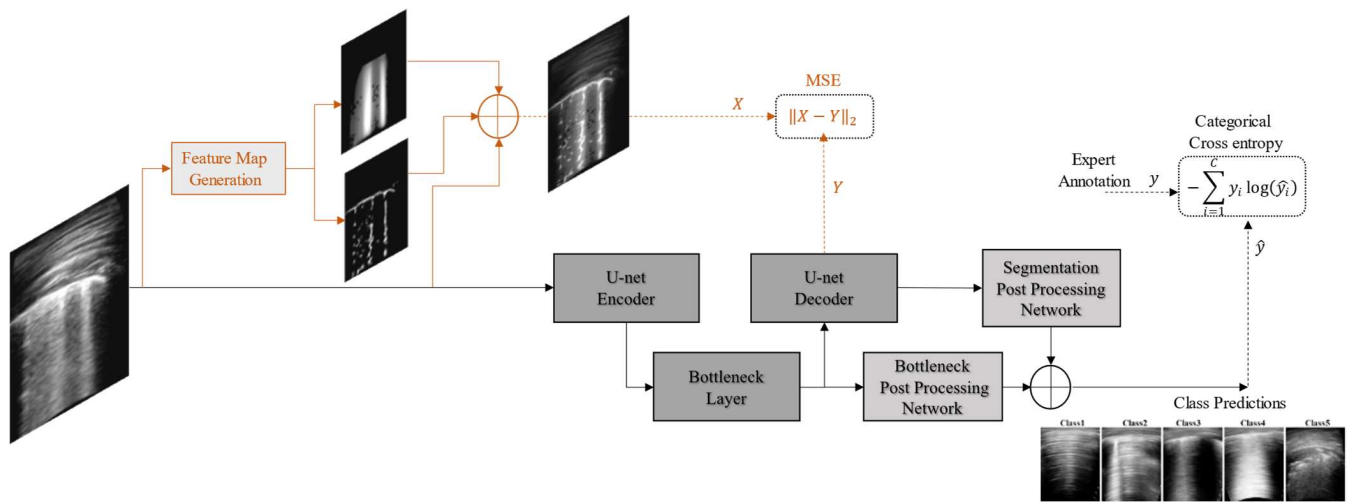

Fig. 5. Proposed LUSNet framework. (Orange lines show the automatic learning of the neural network to focus on the pleural region of interest)

TABLE I
POST PROCESSING NETWORK ARCHITECTURE

|  | Type | Kernel Size | Filters | Activation |
|---|---|---|---|---|
| Segmentation Post Processing Network | Conv2D | 1x1 | 32 | ReLU |
|  | MaxPool2D | 4x4 | - | - |
|  | Batch Normalization | - | - | - |
|  | Conv2D | 1x1 | 32 | ReLU |
|  | MaxPool2D | 4x4 | - | - |
|  | Batch Normalization | - | - | - |
|  | GlobalAveragePooling2D | - | - | - |
| Bottleneck Post Processing Network | GlobalAveragePooling2D | - | - | - |

## III. RESULTS

In this work, about 400 lung ultrasound videos taken from about ten subjects and using different ultrasound machines (Butterfly network, GE, Philips and Fujifilm Sonosite with around 2MHz frequency and Lung preset) and at different sites (Spain and India) are analysed. The videos comprise of healthy lung and various stages of lung consolidation along with the recovery during COVID-19 infection in ten subjects. The data acquisition points are anterior, lateral and PLAPS locations on both left and right sides as per the protocol in [1] resulting in a total of six acquisition points. From the 400 videos, about 5000 randomly chosen image frames over several days from infection to recovery acquired by using three different ultrasound systems are used for the training and validation of the LUSNet. A fair class distribution (1000 images per class) is done to avoid any class imbalance. The 5000 images are divided into five folds of 1000 images (which means 200 images for each class in a fold). Out of the five folds, four folds are employed for training and one-fold is employed for validation. This is repeated by re-initializing the weights of the LUSNet and results are averaged for a five-fold cross validation. For testing, another unseen 1000 images from the same three ultrasound acquisitions are used. To show the importance of the proposed approach to an entirely different data acquisition system, 1000 images from a fourth ultrasound system is also employed.

### A. Reducing Bias in Labelling

As discussed, it is interesting to note that, the fused image itself aid the clinicians for a quick triage in emergency scenarios. This is validated by comparing annotations on about 1000 images by three experts. A similarity ratio is measured as in (5):

$$Similarity\ Score = \frac{Number\ of\ Similar\ Annotations}{1000} \quad (5)$$

For annotations to be considered similar, at least two of the three annotations should be the same. The similarity score achieved for the original images as in the first column of Fig. 4 was around 82% and that of the fused images as in last column of Fig. 4 was higher at 94%. This clearly shows the significance of the proposed acoustic propagation feature maps in aiding the labelling process and potential of the same in aiding the clinicians for a faster triage.

TABLE II
PERFORMANCE COMPARISON OF PROPOSED APPROACH FOR VARIOUS ABLATION STUDIES

| Input Type / Architecture | Class | Backbone: Resnet34 | | | Backbone: VGG16 | | | Backbone: InceptionV3 | | |
|---|---|---|---|---|---|---|---|---|---|---|
|  |  | ACC | SEN | SPEC | ACC | SEN | SPEC | ACC | SEN | SPEC |
| Encoder only with image as input | 1 | 0.98±0.02 | 0.95 | 0.99 | 0.99±0.02 | 0.98 | 0.99 | 0.99±0.01 | 0.98 | 1.00 |
|  | 2 | 0.68±06 | 0.19 | 0.80 | 0.63±0.07 | 0.03 | 0.78 | 0.63±0.07 | 0.03 | 0.78 |
|  | 3 | 0.62±07 | 0.16 | 0.74 | 0.62±0.07 | 0.05 | 0.76 | 0.62±0.07 | 0.05 | 0.77 |
|  | 4 | 0.65±07 | 0.01 | 0.81 | 0.60±0.07 | 0.03 | 0.74 | 0.60±0.07 | 0.04 | 0.74 |
|  | 5 | 0.99±0.01 | 1.00 | 0.99 | 1.00±0 | 1.00 | 1.00 | 1.00±0 | 1.00 | 1.00 |
| Encoder only with multichannel as input | 1 | 0.99±0.01 | 0.99 | 0.99 | 0.98±0.02 | 0.92 | 1.00 | 0.99±0.01 | 0.97 | 1.00 |
|  | 2 | 0.64±0.07 | 0.03 | 0.79 | 0.65±0.07 | 0.13 | 0.78 | 0.63±0.07 | 0.03 | 0.78 |
|  | 3 | 0.64±0.07 | 0.07 | 0.79 | 0.61±0.07 | 0.16 | 0.72 | 0.63±0.07 | 0.10 | 0.77 |
|  | 4 | 0.59±0.07 | 0.08 | 0.72 | 0.65±0.07 | 0.01 | 0.81 | 0.60±± | 0.03 | 0.74 |
|  | 5 | 1.00±0 | 1.00 | 1.00 | 1.00±0.01 | 0.99 | 1.00 | 1.00±0 | 1.00 | 1.00 |
| U-net with image as input | 1 | 0.99±0.01 | 0.98 | 1.00 | 0.98±0.02 | 0.90 | 1.00 | **0.99**±0.01 | **0.99** | **1.00** |
|  | 2 | 0.94±0.03 | 0.85 | 0.97 | 0.92±0.04 | 0.84 | 0.95 | **0.95**±0.03 | **0.87** | **0.96** |
|  | 3 | 0.96±0.03 | 0.89 | 0.98 | 0.96±0.03 | 0.88 | 0.98 | **0.97**±0.03 | **0.87** | **0.99** |
|  | 4 | 0.96±0.03 | 0.93 | 0.97 | 0.95±0.03 | 0.93 | 0.96 | **0.96**±0.03 | **0.93** | **0.97** |
|  | 5 | 1.00±0 | 1.00 | 1.00 | 1.00±0 | 1.00 | 1.00 | **1.00**±0 | **1.00** | **1.00** |
| U-net with multichannel as input | 1 | 0.99±0.01 | 0.99 | 0.99 | 0.99±0.02 | 0.96 | 1.00 | 0.99±0.01 | 0.98 | 1.00 |
|  | 2 | 0.94±0.03 | 0.82 | 0.97 | 0.93±0.04 | 0.84 | 0.95 | 0.94±0.03 | 0.86 | 0.96 |
|  | 3 | 0.95±0.03 | 0.84 | 0.98 | 0.95±0.03 | 0.79 | 0.99 | 0.96±0.03 | 0.88 | 0.98 |
|  | 4 | 0.95±0.03 | 0.92 | 0.96 | 0.95±0.03 | 0.95 | 0.95 | 0.96±0.03 | 0.92 | 0.97 |
|  | 5 | 1.00±0.01 | 1.00 | 1.00 | 1.00±0 | 1.00 | 1.00 | 1.00±0 | 1.00 | 1.00 |
| U-net with fused image as input | 1 | 0.99±0.01 | 0.99 | 0.99 | 0.99±0.02 | 0.98 | 1.00 | 0.99±0.01 | 0.98 | 0.99 |
|  | 2 | 0.94±0.03 | 0.82 | 0.97 | 0.95±0.03 | 0.85 | 0.97 | 0.93±0.03 | 0.83 | 0.96 |
|  | 3 | 0.95±0.03 | 0.84 | 0.98 | 0.96±0.03 | 0.88 | 0.98 | 0.95±0.03 | 0.86 | 0.98 |
|  | 4 | 0.95±0.03 | 0.92 | 0.96 | 0.96±0.03 | 0.93 | 0.96 | 0.95±0.03 | 0.91 | 0.97 |
|  | 5 | 1.00±0.01 | 1.00 | 1.00 | 1.00±0 | 1.00 | 1.00 | 1.00±0 | 1.00 | 1.00 |

TABLE III
PERFORMANCE COMPARISON OF PROPOSED APPROACH FOR A NEW
ULTRASOUND ACQUISITION

| Input Type / Architecture | Class | Backbone: InceptionV3 | | |
|---|---|---|---|---|
| | | ACC | SEN | SPEC |
| Encoder only with image as input | 1 | 0.77±0.06 | 0.10 | 1.00 |
| | 2 | 0.79±0.06 | 0.00 | 0.98 |
| | 3 | 0.25±0.06 | 0.93 | 0.06 |
| | 4 | 0.85±0.05 | 0.00 | 1.00 |
| | 5 | 0.81±0.05 | 0.01 | 0.98 |
| Encoder only with multichannel as input | 1 | 0.82±0.05 | 0.50 | 0.92 |
| | 2 | 0.56±0.07 | 0.23 | 0.63 |
| | 3 | 0.44±0.07 | 0.24 | 0.51 |
| | 4 | 0.87±0.05 | 0.00 | 1.00 |
| | 5 | 0.81±0.05 | 0.08 | 0.96 |
| U-net with image as input | 1 | **0.85±0.05** | **0.51** | **0.97** |
| | 2 | **0.77±0.06** | **0.16** | **0.92** |
| | 3 | **0.72±0.06** | **0.71** | **0.72** |
| | 4 | **0.89±0.04** | **0.30** | **0.99** |
| | 5 | **0.75±0.06** | **0.73** | **0.76** |
| U-net with multichannel as input | 1 | 0.62±0.07 | 0.71 | 0.59 |
| | 2 | 0.75±0.06 | 0.04 | 0.89 |
| | 3 | 0.79±0.06 | 0.41 | 0.91 |
| | 4 | 0.86±0.05 | 0.02 | 1.00 |
| | 5 | 0.68±0.06 | 0.30 | 0.76 |
| U-net with fused image as input | 1 | 0.85±0.05 | 0.57 | 0.94 |
| | 2 | 0.80±0.06 | 0.03 | 1.00 |
| | 3 | 0.71±0.06 | 0.39 | 0.82 |
| | 4 | 0.91±0.04 | 0.16 | 1.00 |
| | 5 | 0.61±0.07 | 0.94 | 0.53 |

*B. Analysis of the proposed approach*

The performance of the proposed approach is compared for various backbones of the U-net architecture such as Resnet34, VGG16 and InceptionV3. Ablations studies with employing only the U-net encoder of LUSNet, which is equivalent to a direct classifier employing one of the backbones, has also been done. The input to both the full LUSNet architecture and the U-net encoder only architecture was also varied between images (the gray scale image is replicated over three channels), multichannel image (where gray scale image, LPI and the SHIBS formed the three channels respectively) and the fused image (where the fused image, which is the sum of gray scale image, LPI and the SHIBS, is replicated over the three channels). The results for this comprehensive comparison are as shown in Table II. The performance of the encoder only architectures are very poor as evident from the results, particularly for classes 2 to 4. Among all the results, the LUSNet with image as input and inceptionV3 as backbone gives the best results and is highlighted by bold blue font in Table II. This opens up the simplest approach compared to multi-channel input or the fused image, as there is no need for calculating the features from the images once the model is fully trained. At the same time, an image similar to the fused image will be available at the output of the U-net decoder as a byproduct. The 95% confidence interval for the accuracy range ($ACC_{95}$) is calculated as per [27] employing (6) is also shown against the accuracy.

$$ACC_{95} = ACC \pm 1.96\sqrt{\frac{ACC(1-ACC)}{200}}$$

Where $ACC$ is the non-percentage test accuracy for 200 images

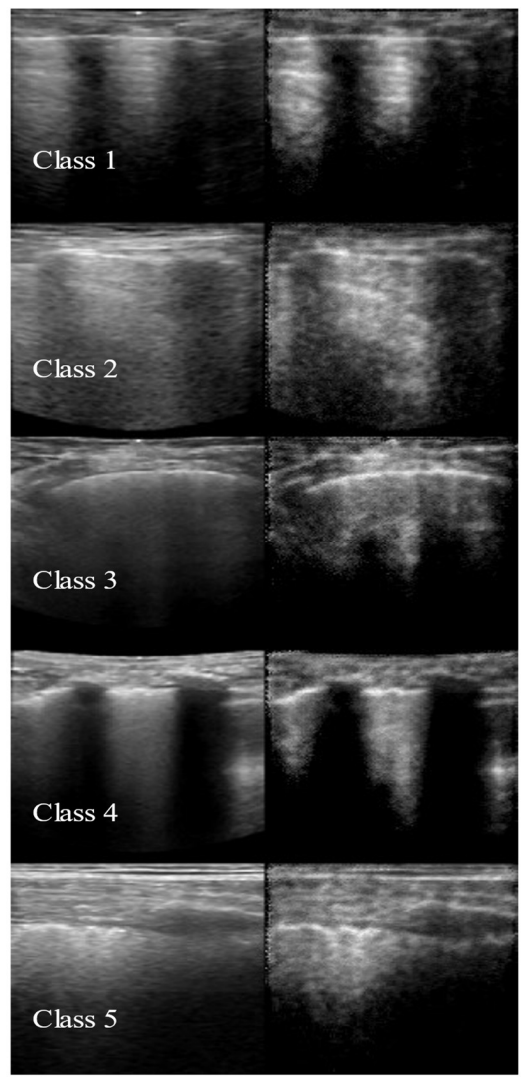

Fig. 6. Sample Enhanced Images for the five classes as obtained at the output of U-net decoder (second column) with images as input to the LUSNet (first column)

per class. To show the agnostic nature of the proposed LUSNet to ultrasound acquisition system, a performance comparison is done for inceptionV3 backbone for the various cases as in Table III. In this case, a new unseen dataset acquired using an entirely new ultrasound system is employed. As expected, the results are in general poor in all the cases, however the proposed LUSNet with images as the input again seems to give the best result and shows the close to agnostic nature of the proposed LUSNet.

Another important aspect of the proposed LUSNet is that the U-net decoder output gives an image which is very close to the fused image and could help the clinicians in a faster triage. Fig. 6 shows the results of some of the sample images and the respective LUSNet U-net decoder output for some of the example cases in Table III. The results clearly show that the LUSNet is learning the pleural region of interest as can be seen from the visible enhancement of the region.

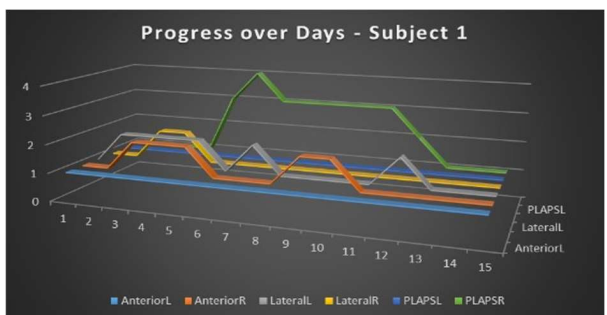

Fig. 7 (a) Scores as predicted by LUSNet (with 97% sensitivity and 96% specificity) for subject 1 over 15 days

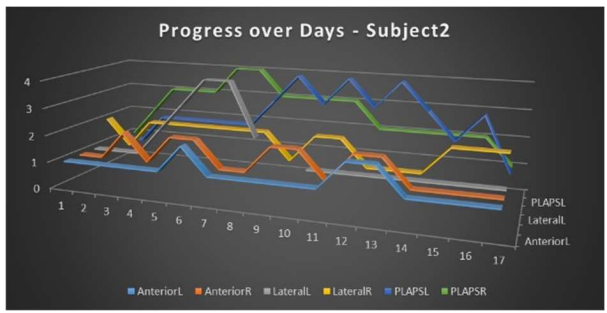

Fig. 7 (b) Scores as predicted by LUSImageNet (with 98% sensitivity and 99% specificity) for subject 2 over 21 days

## IV. Analysis Of COVID-19 Images Over Different Stages

In this section, LUS scans of two COVID-19 positive subjects over fifteen and twenty-one days are analysed using the proposed approach. As discussed, a total of six locations denoted as anteriorL, lateralL, PLAPSL, anteriorR, lateralR, and PLAPSR are tracked for the above days. The scores as predicted by the LUSNet for the above 12 locations for fifteen and twenty-one days are plotted in Fig. 7. The x-axis in Fig. 7 is the number of days, the y-axis is the class predicted and the z-axis is the region of scanning. The sensitivity and specificity of the proposed LUSNet are estimated to be over 95% in both cases. It can be seen that in general, infection has been less mild in subject 1 compared to subject 2 and also in both cases never reached the Class5 stage. In both cases, the right PLAPS and right Anterior regions are with increased severity of infection compared to other regions. This proves that LUS and the scoring system proposed in this paper is capable of tracking the progression of COVID-19 and would be extremely useful to the clinicians for treating ARDS. A detailed analysis of the lung ultrasound scans over six days (days 1, 5, 8, 11, 15, and 21) is presented in this section for the PLAPSR region in subject 2. The LUS images along with the fused image are shown for the above mentioned six days in Fig. 8. As can be seen from LUS images in Fig. 8, the progression from good visibility of A-lines (row 1), pleura consolidation and focused B-lines (row 2), widening of consolidation or multiple B-lines (row 3), improvements in less diverging B-lines (row 4), further reduction in B-line thickness with reappearance of A-lines (row 5) and full reappearance of A-lines (row 6) completes a cycle of full lung recovery in COVID-19 scenario of over twenty-one

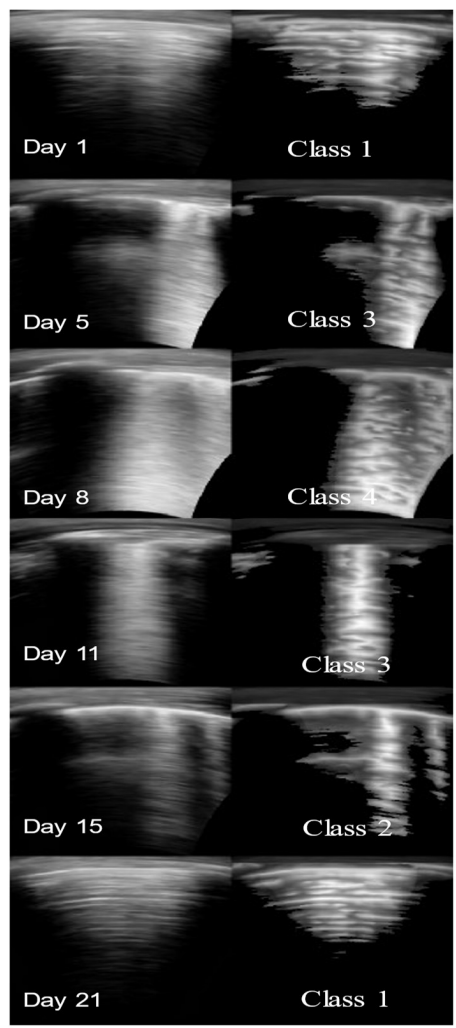

Fig. 8. Random sampling of right PLAPS point LUS images of subject 3 for days 1, 5, 8, 11, 15 and 21 with scores predicted by LUSNet. First column are the original images and the second column are the fused images

days. The second column in Fig. 8, shows the fused images which clearly highlights the pleural region of interest.

## V. Conclusions And Discussion

With the recent outbreak of COVID-19, LUS is fast becoming an inevitable diagnostic tool for continuous monitoring of the lung in the case of ARDS. An important landmark in B-mode US imaging of the lung is the pleura. Analysing the pleura and the subsequent detection of A, B, and C lines will enable quick clinical decisions, particularly when assisting the critically ill. In this work, a novel approach is presented, where acoustic propagation driven features are generated such as local phase image, shadow image and integrated back scatter image are employed to generate a fused image which clearly highlights the pleural region of interest. This fused image is used to train the U-net architecture of the proposed neural network, LUSNet, to learn the pleural region of interest in an unsupervised manner, whereas the LUSNet is also simultaneously trained in a supervised manner to classify the images into five classes of increasing severity of infection.

A detailed analysis of the features shows that the features individually are capable of locating and identifying the presence of A, B, or C lines. The proposed approach has been used for a detailed analysis of LUS of two subjects over infection to full recovery period of COVID-19. Through this work, it has been shown that, it is possible to highlight the pleural region of interest, and further train a neural network to classify the images into various classes, while ensuring that the neural network is learning the pleural region of interest. Even though, in patients with COVID-19 pneumonia, LUS showed a typical bilateral pattern of diffuse interstitial lung syndrome, characterized by multiple or confluent B-lines with spared areas, thickening of the pleural line with pleural line irregularity, and less frequently subpleural consolidations and pleural effusion, it has to be noted that, a full-fledged automatic detection of A-lines and B-lines should consider other lung abnormalities into consideration. This is because, in addition to pulmonary congestion, B-lines are visible in pulmonary fibrosis, granulomatous lung interstitial diseases, atelectasis, lymphangitis, lung contusion, cardiac failure, and ARDS. Similarly, although A-lines are always present in the LUS examinations of the younger subjects (less than 50 years), the majority (94%) of healthy, non-smoker elderly subjects (mean age 79 years) showed no A-lines as reported in [26]. Future work, therefore should consider the above aspects into consideration.

## APPENDIX AND THE USE OF SUPPLEMENTAL FILES

The proposed approach has been deployed for clinical usage and trials at http://www.pulseecho.in/alus/ and codes will be made freely available at https://github.com/maheshpanickeriitpkd/ALUS. The dataset will be made available upon contacting the authors.

## ACKNOWLEDGMENT

The authors would also like to thank the Department of Science and Technology - Science and Engineering Research Board (DSTSERB (CVD/2020/000221)) for the CRG COVID19 funding and the corporate social responsibility (CSR) funding from Federal bank, India.